\documentclass[aps,prd,twocolumns,eqsecnum,preprintnumbers,showpacs,amsmath,amssymb]
{revtex4}

\usepackage{graphicx}% Include figure files
\usepackage{dcolumn}% Align table columns on decimal point
\usepackage{bm}% bold math

\begin{document}

\title{A GENERAL SOLUTION FOR THE QUARK PROPAGATOR IN TWO-DIMENSIONAL COVARIANT GAUGE QCD}

\author{V. Gogohia}
\email[]{gogohia@rmki.kfki.hu}
\author{Gy. Kluge}
\email[]{kluge@rmki.kfki.hu}
\author{I. Vargas de Usera}
\email[]{ignacio@nextlimit.com}

\affiliation{HAS, CRIP, RMKI, Depart.
Theor. Phys., Budapest 114, P.O.B. 49, H-1525, Hungary}

\date{\today}

\begin{abstract}
We have investigated a closed set of equations for the quark
propagator, which has been obtained earlier within a new,
nonperturbative approach to two-dimensional covariant gauge QCD.
It is shown that this theory implies quark confinement (the quark
propagator has no poles, indeed), as well as dynamical breakdown
of chiral symmetry (a chiral symmetry preserving solution is
forbidden). The above-mentioned set of equations can be exactly
solved in the chiral limit.
 We develop an analytical formalism, the so-called chiral
perturbation theory at the fundamental quark level, which allows
one to find solution for the quark propagator in powers of the
light quark masses. Each correction satisfies the differential
equation, which can be formally solved.
 We develop also an analytical formalism which allows one to find solution
 for the quark propagator in the inverse powers of the heavy quark masses.
It coincides with the free heavy quark propagator up to terms of
order $1 / m_Q^3$, where $m_Q$ is the heavy quark mass. So this
solution automatically possesses the heavy quark flavor symmetry
up to terms of order $1 / m_Q$. At the same time, we have found a
general solution for the heavy quark propagator, which by no means
can be reduced to the free one.
\end{abstract}



\maketitle

\section{Introduction}

The investigation of two-dimensional (2D) QCD in the context of
the Schwinger-Dyson (SD) dynamical equations of motion has been
initiated by the pioneering paper of 't Hooft \cite{1}. He used
the free gluon propagator in the light-cone gauge, which is free
from ghost complications. He used also the large $N_c$ (the number
of colors) limit technique in order to make the perturbation (PT)
expansion with respect to $1/N_c$ reasonable. In this case the
planar diagrams are reduced to quark self-energy and ladder
diagrams, which can be summed. The bound-state problem within the
Bethe-Salpeter (BS) formalism was finally obtained free from the
infrared (IR) singularities. The existence of a discrete spectrum
only (no continuum in the spectrum) was demonstrated. Since this
pioneering paper 2D QCD continues to attract attention (see, for
example, review \cite{2} and recent papers \cite{3,4,5} and
references therein). Despite its simplistic vacuum structure it
remains a rather good laboratory for the modern theory of strong
interactions, which is four-dimensional (4D) QCD \cite{6}.

In our previous publications \cite{7,8} we have investigated 2D
QCD in the arbitrary covariant gauge for the first time. In these
works a new, nonperturbative (NP) solution (using neither large
$N_c$ limit technique explicitly nor a weak coupling regime, i.e.,
ladder approximation) to 2D QCD in the covariant gauge is obtained
in the context of the above-mentioned SD equations, complemented
by the corresponding Slavnov-Taylor (ST) identities. It is well
known, however, that covariant gauges, in general, are complicated
by the ghost contributions. Nevertheless, we have shown that ghost
degrees of freedom can be considerable within our approach
\cite{7}. The ghost-quark sector contains a very important piece
of information on quark degrees of freedom themselves through the
corresponding quark ST identity. This is just the information
which should be self-consistently taken into account. In this way
a close set of equations has been derived for the quark propagator
\cite{7}. The main purpose of this Letter is to exactly solve the
obtained system of equations in the chiral limit and to develop
analytical methods of its solution in the general case, i.e., for
the nonzero current quark masses. Let us emphasize in advance that
we have found a general solution for the heavy quark propagator,
which by no means could be reduced to the free one. All of this
will provide the necessary basis for future numerical calculations
as well.

\section{Quark SD equation}

The final system of equations, obtained in Ref. \cite{7} for the
quantities in the quark sector, are presented by the quark SD
equation and the quark ST identity as follows (Euclidean
signature):

\begin{eqnarray}
S^{-1} (p) &=& S_0^{-1} (p)+ \bar g^2 \Gamma_\mu(p,0) S(p)
\gamma_\mu,
\nonumber \\
\Gamma_\mu(p,0) &=& id_\mu S^{-1}(p) - S(p) \Gamma_\mu(p,0)
S^{-1}(p).
\end{eqnarray}
For simplicity, here we remove an overbar from the definitions of
the renormalized Green's functions, retaining it only for the
coupling constant $\bar g$ (which has the dimensions of mass) in
order to distinguish it from initial ("bare") coupling constant.
It contains all known finite numerical factors. $\Gamma_\mu(p,0)$
is obviously the proper quark-gluon vertex at zero momentum
transfer. The Euclidean version of our parametrization of the
quark propagator is as follows: $i S(p) = \hat p A(p^2) - B(p^2)$.
It is convenient to introduce the dimensionless variables and
functions as $A(p^2) = \bar g^{-2} A(x), \ B(p^2) = \bar g^{-1}
B(x), \ x = p^2/{\bar g^2}$. Performing further some tedious
algebra of the $\gamma$ matrices in 2D Euclidean space, the system
(2.1) can be explicitly reduced to a system of a coupled,
nonlinear ordinary differential equations of the first order for
the $A(x)$ and $B(x)$ quark propagator form factors, namely

\begin{eqnarray}
 x A' &=& - (1 + x) A - 1 - \bar m_0 B, \nonumber\\
2B B' &=& - A^2  + 2 ( \bar m_0 A - B)B,
\end{eqnarray}
 where $A \equiv A(x)$, $B \equiv B(x)$, and
the prime denotes the derivative with respect to the Euclidean
dimensionless momentum variable $x$. For the dimensionless current
quark mass, we introduce the notation $\bar m_0 = m_0 / \bar g$.

 The formal exact solution of the system (2.2) for the dynamically generated
quark mass function is

\begin{equation}
 B^2(c, \bar m_0; x) =  \exp(- 2x)
\int^c_x \exp(2x') \tilde{\nu}(x')\, dx' ,
\end{equation}
and $c$ is the constant of integration. Not losing generality, it
can be fixed as $c = p^2_c / \bar g^2$, where $p^2_c$ is some
constant momentum squared, and

\begin{equation}
\tilde{\nu} (x) = A^2(x) +2 A(x) \nu(x)
\end{equation}
with

\begin{equation}
\nu (x) = - \bar m_0 B(x) = xA'(x) + (1 + x)A(x) + 1.
\end{equation}
 Then the equation determining the $A(x)$ function becomes

\begin{equation}
{d \nu^2 (x) \over dx}+ 2\nu^2(x)= - A^2(x) \bar m^2_0 - 2 A(x)
\nu(x) \bar m_0^2.
\end{equation}

\subsection{Quark confinement}

 As was emphasized in Refs.\cite{7,8}, the important observation
is that the formal exact solution (2.3) exhibits the algebraic
branch point at $x=c$, which completely $excludes \ a \ pole-type
\ singularity$ at any finite point on the real axis in the
$x$-complex plane whatever the solution for the $A(x)$ function
might be. Thus the solution for the quark propagator cannot be
presented as the expression having finally a pole-type singularity
at any finite point $p^2 = - m^2$ (Euclidean signature), i.e.,

\begin{equation}
S(p) \neq {const \over \hat p + m},
\end{equation}
certainly satisfies thereby the first necessary condition of quark
confinement, formulated at the fundamental quark level as the
absence of a pole-type singularity in the quark propagator
\cite{9}. It is well known that such kind of unphysical
singularity (algebraic branch point at $x=c$) is due to the
inevitable ghost contributions in the covariant gauge QCD.
However, as was explained in Refs. \cite{7,8}, it will not cause
any problems within our approach in order to calculate truly NP
quantities, such as quark condensate. The absence of a pole-type
singularities in the quark propagator as a criterion of
confinement at the microscopic level is only first necessary
condition. The second sufficient condition of this criterion,
formulated at the macroscopic (hadron) level, is the existence of
a discrete spectrum only (no continuum in the spectrum) in the
bound-state problem within the corresponding BS formalism
\cite{1}. Its discussion is obviously beyond the scope of the
present Letter.

\subsection{Dynamical breakdown of chiral symmetry (DBCS)}

 From a coupled system of the differential equations (2.2) it is
easy to see that this system $allows \ a  \ chiral \ symmetry \
breaking \ solution \ only$,

\begin{equation}
\bar m_0 = 0, \quad A(x) \ne 0, \ B(x) \ne 0
\end{equation}
and $forbids \ a \ chiral \ symmetry \ preserving \ solution$,

\begin{equation}
\bar m_0 = B(x) = 0, \quad A(x) \ne 0.
\end{equation}
Thus any nontrivial solution automatically breaks the $\gamma_5$
invariance of the quark propagator, and therefore $certainly$
leads to the spontaneous chiral symmetry breakdown at the
fundamental quark level ($m_0 = 0, \ \overline{B}(x) \ne 0$,
dynamical quark mass generation). In all previous investigations a
chiral symmetry preserving solution always exists. For simplicity,
we do not distinguish between $B(x)$ and $\overline{B}(x)$,
calling both dynamically generated quark mass functions.

A few remarks are in order. A nonzero, dynamically generated quark
mass function defined by condition (2.8) is the order parameter of
DBCS at the fundamental quark level. At the phenomenological level
the order parameter of DBCS is the nonzero chiral quark condensate
determined as $< \bar q q>_0 \sim - \bar g \int_0^{c_0} dx \
B_0(c_0, x)$ within our approach (see Ref. \cite{8} and  $B_0(c_0,
x)$ is explicitly given below in Eq. (2.11)). In general, it can
be formally zero, even when the mass function is definitely
nonzero. Thus the nonzero, dynamically generated quark mass is a
much more appropriate condition of DBCS than the quark condensate.
One may say that this is the first necessary condition of DBCS,
while the nonzero chiral quark condensate is the second sufficient
one.

\subsection{The chiral limit}

 In the chiral limit ($\bar m_0 = 0$) the system (2.2) can be solved
exactly. The solution for the $A(x)$ function is

\begin{equation}
A_0(x) =  - x^{-1} \left\{ 1 - \exp(-x) \right\}.
\end{equation}

It has thus the correct asymptotic properties (see Fig. 1).
 It is regular at small $x$ and asymptotically approaches the free
propagator at infinity ($x \rightarrow \infty$), which can be
formally achieved by the two ways: $p^2 \rightarrow \infty$ at
fixed $\bar g^2$ and/or by $\bar g^2 \rightarrow 0$ as well. Let
us note that the last limit is known as the PT one. For the
dynamically generated quark mass function $B(x)$ the exact
solution is

\begin{equation}
 B^2_0(c_0, x) =  \exp(- 2x)
\int^{c_0}_x \exp(2x') A^2_0(x')\, dx',
\end{equation}
where $c_0 = p_0^2 / \bar g^2$ is an arbitrary constant of
integration and $p^2_0$ is some constant momentum squared for the
chiral limit case. It is regular at zero. In addition, it also has
algebraic branch points at $x=c_0$ and at infinity (at fixed
$c_0$, i.e., when $\bar g^2$ is fixed). As in the general
(nonchiral) case, these unphysical singularities are caused by the
inevitable ghost contributions in the covariant gauges (for
general behavior of this solution see Fig. 2).

\begin{figure}
\includegraphics[scale=0.6,angle=-90]{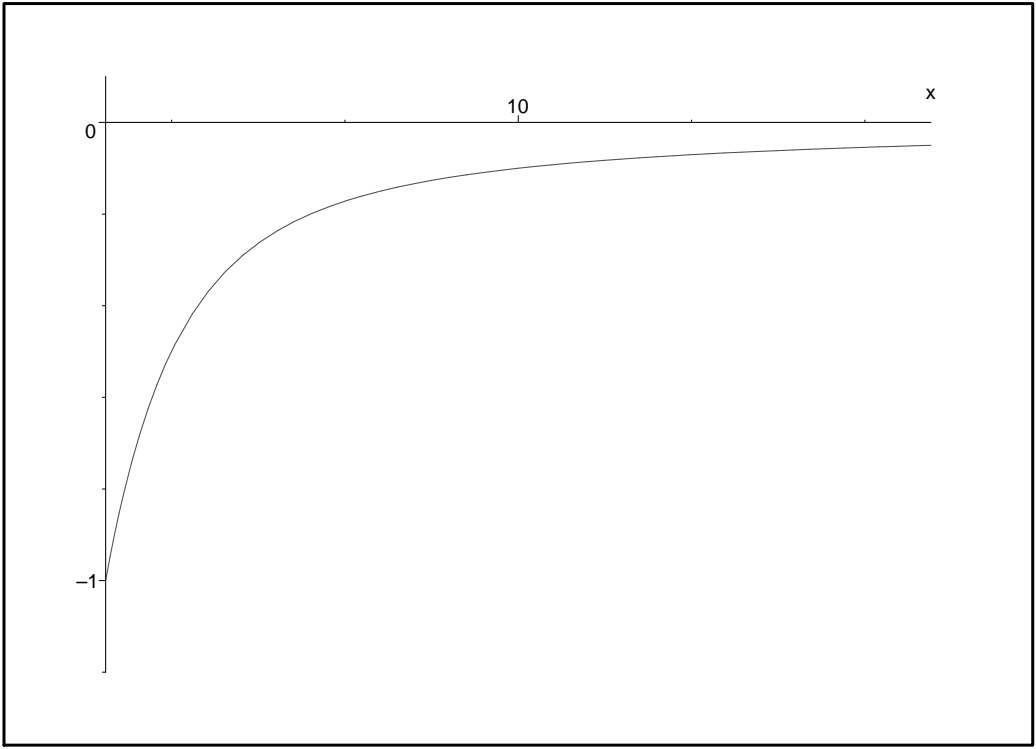}
 \caption{\label{fig1} $A_0(x)$ as given by Eq.(2.10).}
\end{figure}

\begin{figure}
\includegraphics[scale=0.6,angle=-90]{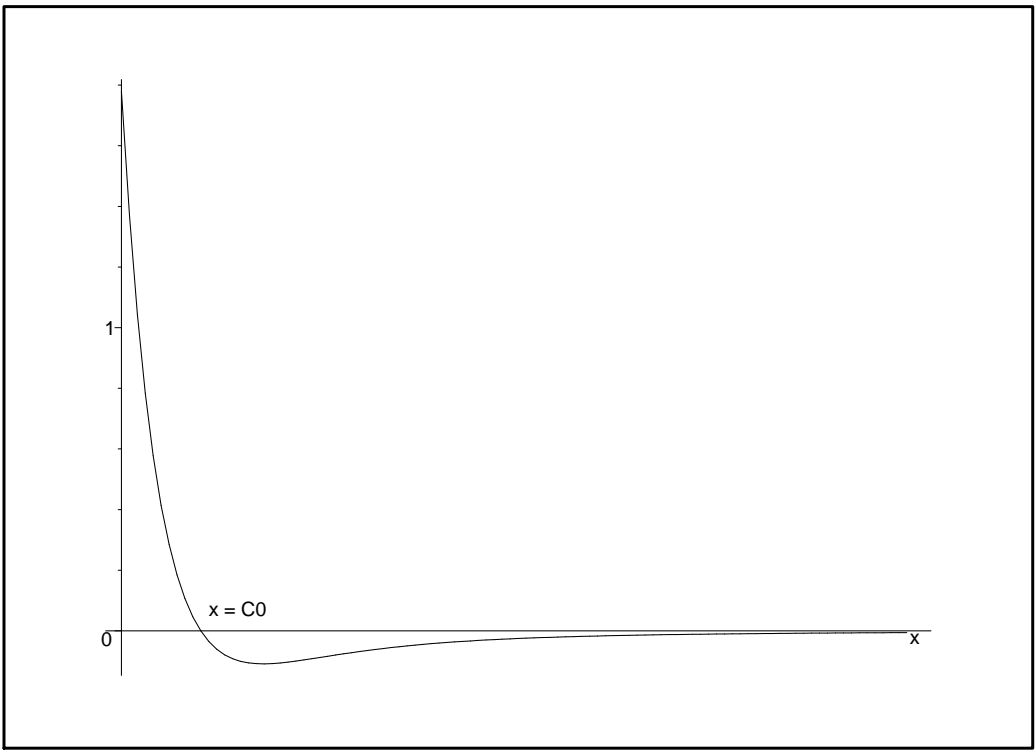}
 \caption{\label{fig2} The dynamically generated quark mass function
  as given by Eq. (2.11).}
\end{figure}

   As was mentioned above, $A_0(x)$ automatically has a correct
behavior at infinity (it does not depend on the constant of
integration, since it was specified in order to get regular at
zero solution). In the PT limit ($\bar g^2 \rightarrow 0$), the
constant of the integration $c_0$ and the variable $x$ go to
infinity uniformly ($c_0, x \rightarrow \infty$), so the
dynamically generated quark mass function (2.11) identically
vanishes in this limit, in accordance with the vanishing current
light quark mass in the chiral limit. Obviously, we have to keep
the constant of integration $c_0$ in Eq. (2.11) arbitrary but
finite in order to obtain a regular at zero point solution. The
problem is that if $c_0 = \infty$, then the solution (2.11) does
not exist at all at any finite $x$, in particular at $x=0$. This
is valid, of course, for the general solution (2.3) as well.

\section{Nonzero current quark masses}

Let us formulate and develop now the calculation scheme, which
gives the solution of the system (2.2) step by step in powers of
the light nonzero current quark masses, as well as in the inverse
powers of the heavy quark masses. For this purpose, it is much
more convenient to start from the ground system itself, Eqs.
(2.2), rather than to investigate the general solution (2.3). For
this purpose, let us rewrite the ground system (2.2) as follows:

\begin{eqnarray}
 x A' + (1 + x) A + 1 &=& - \bar m_0 B, \nonumber\\
2B B' + A^2 + 2 B^2 &=& 2 \bar m_0 A B.
\end{eqnarray}
  As was mentioned above, we are interested in the solutions which
are $regular \ at \ zero$ and asymptotically approach the free
quark case at infinity. Because of our parametrization of the
quark propagator, its asymptotics have to be determined as follows
(Euclidean signature): $A (x) \sim_{x \rightarrow \infty} - 1 / (x
+ \bar m_0^2), \quad B (x) \sim_{x \rightarrow \infty} - \bar m_0
/(x + \bar m_0^2) $, and neglecting $\bar m_0^2$ in the
denominators for light quarks.
 The ground system (3.1) is very suitable for numerical
calculations.

\subsection{Light quarks}

Let us now develop the above-mentioned analytical formalism, which
makes it possible to find solution of the ground system (3.1) step
by step in powers of the light ($u, \ d, \ s$) nonzero current
quark masses, the so-called chiral perturbation theory at the
fundamental quark level. For this purpose it is convenient to
present the quark propagator form factors $A$ and $B$ as follows:

\begin{equation}
A (x) = \sum_{n=0}^{\infty} \bar m_0^n A_n (x), \quad B (x) =
\sum_{n=0}^{\infty} \bar m_0^n B_n (x),
\end{equation}
where it is formally assumed that $\bar m_0^{(u,d,s)} \ll 1$.
Substituting these expansions into the ground system (3.1) and
omitting some tedious algebra, one obtains

\begin{eqnarray}
 x A'_0(x) + (1 + x) A_0(x) + 1 &=& 0 , \nonumber\\
2B_0(x) B'_0(x)+ A^2_0(x) + 2 B^2_0(x) &=& 0,
\end{eqnarray}
and for $n=1,2,3,...$, one obtains

\begin{eqnarray}
x A'_n(x) + (1 + x) A_n(x) &=& - B_{n-1}(x) , \nonumber\\
2P_n(x) + M_n(x) + 2 Q_n(x) &=& 2 N_{n-1}(x),
\end{eqnarray}
where

\begin{eqnarray}
P_n (x) &=& \sum_{m=0}^n B_{n-m} (x) B'_m (x), \quad
M_n (x) = \sum_{m=0}^n A_{n-m} (x) A_m (x), \nonumber\\
Q_n (x) &=& \sum_{m=0}^n B_{n-m} (x) B_m (x), \quad N_n (x) =
\sum_{m=0}^n A_{n-m} (x) B_m (x).
\end{eqnarray}

Is is obvious that the system (3.3) describes the ground system
(3.1) in the chiral limit ($\bar m_0=0$). As we already know, it
can be solved exactly (see below as well). The first nontrivial
correction in powers of small $\bar m_0$ is determined by the
following system, which follows from Eqs. (3.4) and (3.5), and it
is

\begin{equation}
x A'_1 + (1 + x) A_1 = - B_0, \ (B_1B'_0 + B_0 B'_1) + A_0 A_1 + 2
B_0 B_1 = A_0 B_0,
\end{equation}
where we omit the dependence on the argument $x$, for simplicity.
In the similar way can be found the system of equations to
determine terms of order $m_0^2$ in the solution for the quark
propagator and so on.

Let us present a general solution to the first of Eqs. (3.4),
which is

\begin{equation}
 A_n(x) = - x^{-1} e^{-x} \int_0^x dx' \ e^{x'} B_{n-1} (x').
\end{equation}
It is always regular at zero, since all $B_n(x)$ are regular as
well. The advantage of the developed chiral perturbation theory at
the fundamental quark level is that each correction in the powers
of small current quark masses is determined by the corresponding
system of equations which can be formally solved exactly.

Let us write down the system of solutions approximating the light
quark propagator up to first correction, i.e.,

\begin{eqnarray}
A(x) &=& A_0(x) + \bar m_0 A_1(x) + ...., \nonumber\\
B(x) &=& B_0(x) + \bar m_0 B_1(x) + ....
\end{eqnarray}
This system is

\begin{equation}
A_0 (x) = - x^{-1} (1 - e^{-x}), \quad A_0(0) = -1 ,
\end{equation}

\begin{equation}
 B^2_0(x) = e^{-2x} \int_x^{c_0} dx' \  e^{2x'} A_0^2 (x'),
\end{equation}
 and

\begin{equation}
 A_1(x) = - x^{-1} e^{-x} \int_0^x dx' \ e^{x'} B_0 (x'),
\end{equation}

\begin{equation}
 B_1(x) = e^{-2x}B_0^{-1}(x) \int_{c_1}^x dz \ e^{2z} A_0(z)[B_0(z)
 -A_1(z)].
\end{equation}
 In physical applications we also need $B^2(x)$, so we have

\begin{eqnarray}
B^2(x) &=& B_0^2(x) + 2 \bar m_0 B_0(x) B_1(x) + ... \nonumber\\
&=& B_0^2(x) + 2 \bar m_0 e^{-2x} \int_{c_1}^x dz \ e^{2z}
A_0(z)[B_0(z) -A_1(z)]  + ...,
\end{eqnarray}
and the relation between constants of integration $c_0$ and $c_1$
remains, in general, arbitrary. However, there exists a general
restriction, namely $B^2(x) \geq 0$ and it should be real, which
may lead to some bounds for the constants of integration, while $x
\leq c_0$ always remains valid.

\subsection{Heavy quarks}

For heavy quarks ($c, \ b, \ t$) it makes sense to replace $\bar
m_0 \rightarrow \bar m_Q$. In this case it is convenient to find
solution for heavy quark form factors $A$ and $B$ as follows:

\begin{equation}
\bar m_Q^2A (x) = \sum_{n=0}^{\infty} \bar m_Q^{-n} A_n (x), \quad
\bar m_Q B (x) = \sum_{n=0}^{\infty} \bar m_Q^{-n} B_n (x),
\end{equation}
and for heavy quark masses it is formally assumed that $\bar
m_Q^{(c,b,t)} \gg 1$, i.e., the inverse powers are small.
Substituting these expansions into the first equation of the
ground system (3.1) and omitting some tedious algebra, one finally
obtains

\begin{equation}
B_0(x) = -1, \quad B_1(x) = 0,
\end{equation}
and

\begin{equation}
xA'_n(x) + (1 + x)A_n (x) = - B_{n+2}(x), \quad n= 0,1,2,3,...
\end{equation}
In the same way, by equating terms at equal powers in the inverse
of heavy quark masses, from second of the equations of the ground
system (3.1), one finally obtains

\begin{equation}
P_0(x) + Q_0(x) - N_0(x) = 0, \quad P_1(x) + Q_1(x) - N_1(x)= 0,
\end{equation}
and

\begin{equation}
P_{n+2}(x) + Q_{n+2}(x) - N_{n+2}(x) = -{1 \over 2}M_n(x),
  \quad n=0,1,2,3,...,
\end{equation}
where $P_n(z), \ M_n(z), \ Q_n(z), \ N_n(z)$ are again given by
Eqs. (3.5). Solving Eqs. (3.17) and taking into account Eq.
(3.15), one obtains

\begin{equation}
A_0(x) = B_0(x)= -1, \quad A_1(x) = B_1(x) = 0,
\end{equation}
so the final system to be solved further becomes

\begin{eqnarray}
x A'_n(x) + (1+x) A_n(x) &=& - B_{n+2}(x), \nonumber\\
P_{n+2}(x) + Q_{n+2}(x) - N_{n+2}(x) &=& - {1 \over 2} M_n(x),
\quad n = 0,1,2,3, ...
\end{eqnarray}
It is possible to show that all odd terms are simply zero, i.e.,
$A_{2n+1}(x)= B_{2n+1}(x) = 0, \quad n = 0,1,2,3, ...$.

The explicit solutions for a few first nonzero terms are

\begin{equation}
A_0(x)= B_0(x) = - 1.
\end{equation}

\begin{equation}
A_2(x) = x + {3 \over 2}, \quad B_2(x) = x + 1.
\end{equation}

\begin{equation}
A_4(x) = - x^2 - {3 \over 2} x - {15 \over 2}, \quad B_4(x) = -
x^2 - {7 \over 2} x - {3 \over 2}.
\end{equation}
Thus our solutions for the heavy quark form factors look like

\begin{eqnarray}
A (x) &=& { 1 \over \bar m_Q^2} \sum_{n=0}^{\infty} \bar m_Q^{-n} A_n (x) \nonumber\\
 &=& - {1 \over \bar m_Q^2} + { x \over \bar m_Q^4} - {x^2 \over \bar m_Q^6} + ...+  D_A (x),
\end{eqnarray}
where

\begin{equation}
D_A(x) = { 3 \over 2 \bar m_Q^4} - { 3x + 15 \over 2 \bar m_Q^6} +
...,
\end{equation}
and

\begin{eqnarray}
B(x) &=& { 1 \over \bar m_Q} \sum_{n=0}^{\infty} \bar m_Q^{-n} B_n (x) \nonumber\\
 &=& - {1 \over \bar m_Q} + { x \over \bar m_Q^3} - {x^2 \over \bar m_Q^5} + ...+  D_B (x),
\end{eqnarray}
where

\begin{equation}
D_B(x) = { 1 \over  \bar m_Q^3} - {7 x + 3 \over 2 \bar m_Q^5} +
...
\end{equation}
Summing up, one obtains

\begin{eqnarray}
A (x) &=& - { 1 \over x + \bar m_Q^2} + D_A (x), \nonumber\\
B(x) &=& - {m_Q \over x+ \bar m_Q^2} +  D_B (x).
\end{eqnarray}

In terms of the Euclidean dimensionless variables, the quark
propagator is

\begin{equation}
iS(x) = \hat x A(x) - B(x).
\end{equation}
 Using our solutions, obtained above, it can be written down as
follows:

\begin{equation}
iS_h(x) = iS_0(x) + \hat x D_A(x)  - D_B(x),
\end{equation}
where $iS_0(x)$ is nothing else but the free quark propagator with
the substitution $\bar m_0 \rightarrow \bar m_Q$, i.e.,

\begin{equation}
iS_0(x) = - { \hat x - \bar m_Q \over x + \bar m_Q^2}.
\end{equation}
Since $\hat x D_A(x)-D_B(x)$ is of order $\bar m_Q^{-3}$, then Eq.
(3.30), becomes

\begin{equation}
iS_h(x) = iS_0(x) + O(\bar m_Q^{-3}),
\end{equation}
i.e., it is reduced to the free quark propagator up to terms of
order $1 / \bar m_Q^3$.

\subsection{Heavy quarks flavor symmetry}

It is instructive to show explicitly that our solution (3.32)
possesses the heavy quark flavor symmetry \cite{10,11}. We will
show that the quark propagator to leading order (up to terms of
order $1 / m_Q$) in the inverse powers of the heavy quark mass
will not depend on it, i.e., it is a manifestly flavor independent
to the leading order of this expansion. For this purpose, we must
take into account that the argument $x$, which is the
dimensionless momentum of the heavy quark, contains itself the
heavy quark mass $m_Q$. In other words, a standard heavy quark
momentum decomposition should be used, namely

\begin{equation}
p_{\mu} = m_Q \upsilon_{\mu} + k_{\mu},
\end{equation}
as well as

\begin{equation}
\hat x = \gamma_{\mu} x_{\mu} = \gamma_{\mu} (m_Q \upsilon_{\mu} +
y_{\mu}),
\end{equation}
where $\upsilon$ is the four-velocity with $\upsilon^2=-1$
(Euclidean signature). It should be identified with the
four-velocity of the hadron. The "residual" momentum $k$ is of
dynamical origin. In these terms the Euclidean dimensionless
dynamical momentum variable $x=p^2/ \bar g^2$ then becomes

\begin{equation}
x = - \bar m_Q^2 - 2 \bar m_Q t - z,
\end{equation}
where we denote $t = (\upsilon \cdot y)$ with $y_{\mu} = k_{\mu} /
\bar g$ and $z= k^2/ \bar g^2$.

 Substituting these expressions into the Eq. (3.31) and taking into account
only the leading order term in the inverse powers of $m_Q$, one
finally obtains

\begin{equation}
iS_h(\upsilon, y) = iS_0(\upsilon, y) + O({1 \over m_Q}),
\end{equation}
where

\begin{equation}
iS_0(\upsilon, y) = {1 \over \upsilon \cdot y } {\hat \upsilon  -
1 \over 2},
\end{equation}
which is exactly the heavy quark propagator \cite{11}. Thus our
propagator does not depend on $m_Q$ to leading order in the heavy
quark mass limit, $m_Q \rightarrow \infty$, i.e., in this limit it
possesses the heavy quark flavor symmetry, indeed.

\section{The general solution for heavy quarks}

It is easy to understand that the chiral perturbation theory at
the fundamental quark level developed for light quarks in
subsection A completely coincides with the general solution (2.3),
complemented by Eqs. (2.4), (2.5) and (2.6). We use these
equations in the chiral limit as input information in the
expansions (3.2). However, things are not so straightforward in
the case of heavy quarks. Developing the chiral perturbation
theory in the inverse powers of the heavy quark masses in
subsection B, we do not use the general solution (2.3), only the
system (2.2) itself. In this subsection we will show that the
general solution (2.3), complemented by Eqs. (2.4), (2.5) and
(2.6), for the heavy quark mass function possesses much more
information than the direct solution of the system (2.2) on
account of the expansions (3.14) provides at all.

Starting from the expansion (3.14) for the $A(x)$ function, which
contributes into the quark wave function renormalization only, and
using exact Eqs. (2.5) and (2.6), it is possible to show
explicitly that it is determined by the solution (3.28), i.e., it
is

\begin{equation}
A (x) = - { 1 \over x + \bar m_Q^2} + D_A (x),
\end{equation}
where the $D_A(x)$ function is given in Eq. (3.25). Thus in the
case of the $A(x)$ function the straightforward solution of the
initial system (2.2) completely coincides with exact solution,
indeed.

Unfortunately, things are not so simple for the heavy quark mass
function $B(x)$, which should be determined with the help of the
exact solution (2.3), on account of the solution (4.1).
Substituting it first into the relation (2.5) and then using the
relation (2.4) and doing some tedious algebra, one finally obtains

\begin{equation}
\tilde{\nu}(x) = { 1 - 2 \bar m_Q^2 \over (x + \bar m_Q^2)^2} + {2
\bar m_Q^2 \over (x + \bar m_Q^2)^3} + \bar D_A (x),
\end{equation}
where

\begin{eqnarray}
\bar D_A (x) &=& D_A^2(x) + 2 D_A(x) - {2 x \over (x + \bar
m_Q^2)} D'_A(x)  + {2 x \over (x +
\bar m_Q^2)^2} D_A(x) \nonumber\\
&+& 2x D_A(x)D'_A(x) - {2(3+ 2x) \over (x + \bar m_Q^2)} D_A(x) +
2(1+x) D^2_A(x).
\end{eqnarray}

So the general solution (2.3) becomes

\begin{eqnarray}
 B^2(c, \bar m_Q; x) &=& (1 - 2 \bar m_Q^2) \exp(- 2x)
\int^c_x { \exp(2x') \over (x' + \bar m_Q^2)^2} dx' \nonumber\\
&+& 2 \bar m_Q^2 \exp(- 2x) \int^c_x { \exp(2x') \over (x' + \bar
m_Q^2)^3} dx' + \bar D_B(c, \bar m_Q;x),
\end{eqnarray}
where

\begin{equation}
\bar D_B(c, \bar m_Q; x) = \exp(- 2x) \int^c_x \exp(2x') \bar
D_A(x') dx'.
\end{equation}
The first two integrals can be explicitly expressed in terms of
the corresponding integral exponential function $Ei(x)$, so one
has

\begin{eqnarray}
 B^2(c, \bar m_Q; x) &=& (1 - 2 \bar m_Q^2) e^{-2x}
\Bigl[ {e^{2x} \over (x+ \bar m_Q^2)} - {e^{2c} \over (c+ \bar
m_Q^2)} - 2e^{-2\bar m_Q^2}[
Ei(2(x+\bar m_Q^2))- Ei(2(c+\bar m_Q^2))] \Bigr] \nonumber\\
&+& 2 \bar m_Q^2 e^{-2x} \Bigl[ -{e^{2c} \over 2(c+\bar
m_Q^2)^2} - {e^{2c} \over (c+\bar m_Q^2)} + 2e^{-2\bar m_Q^2} Ei(2(c+\bar m_Q^2)) \nonumber\\
&+&  {e^{2x} \over 2(x+\bar m_Q^2)^2} + {e^{2x} \over (x+\bar
m_Q^2)} - 2e^{-2\bar m_Q^2} Ei(2(x+\bar m_Q^2)) \Bigr] + \bar
D_B(c, \bar m_Q;x).
\end{eqnarray}
It is convenient to separate the dependence on the constant of
integration $c$, so after some algebra one obtains

\begin{eqnarray}
B^2(c, \bar m_Q; x) &=& {1 \over (x+\bar m_Q^2)} + {\bar m_Q^2
\over (x+\bar m_Q^2)^2}
- 2e^{-2(x+ \bar m_Q^2)} Ei(2(x+\bar m_Q^2)) \nonumber\\
&+& \bar D_B(c, \bar m_Q;x)+ \tilde{D}_B(c, \bar m_Q;x),
\end{eqnarray}
where

\begin{eqnarray}
\tilde{D}_B(c, \bar m_Q; x) &=& - {1 - 2 \bar m_Q^2 \over (c+ \bar
m^2_Q)}e^{2(c-x)} +
2e^{-2(x+ \bar m_Q^2)} Ei(2(c+\bar m_Q^2)) \nonumber\\
&-& 2 \bar m_Q^2  \Bigl[ {1 \over 2(c+\bar m_Q^2)^2} + {1 \over (c
+ \bar m_Q^2)} \Bigl]e^{2(c-x)}.
\end{eqnarray}
Evidently, by no means the exact solution (4.7) can be reduced to
the free quark propagator solution (3.28) for the $B(x)$ function,
except, maybe, of the asymptotical regime (se below). It is
regular at zero.

It identically vanishes in the PT limit $\bar g^2 \rightarrow 0$,
as it should be, since in this case $c=x \rightarrow \infty$
uniformly (see general solution (2.3)). In the heavy quark mass
infinite limit ($m_Q \rightarrow \infty$ and $\bar g^2$ fixed),
the quark momentum also goes to infinity (i.e., $x \rightarrow
\infty$ see, for example Eq. (3.33)). In this case the constant of
integration $c$ remains finite, and therefore the composition
$\tilde{D}_B(c, \bar m_Q; x)$ also vanishes. Using further the
asymptotics of the integral exponential function $Ei(z)$ as
follows:

\begin{equation}
Ei(z)) = e^z \Bigr[ {1 \over z} +   {1 \over z^2} +O({1 \over
z^3}) \Bigl], \quad z \rightarrow \infty,
\end{equation}
from Eq. (4.7) one obtains

\begin{equation}
B^2(c, \bar m_Q; x) \sim_{x, \bar m_Q^2 \rightarrow \infty} {\bar
m_Q^2 \over (x+\bar m_Q^2)^2} + \bar D_B(c, \bar m_Q;x).
\end{equation}
Choosing negative sign in the square root, one finally obtains

\begin{equation}
B(\bar m_Q; x) \sim_{x, \bar m_Q^2 \rightarrow \infty} - {\bar m_Q
\over (x+\bar m_Q^2)} \Bigl[ 1 + {(x + \bar m^2_Q)^2 \over \bar
m_Q^2} \bar D_B(\bar m_Q;x) \Bigr]^{1 \over 2},
\end{equation}
where $\bar D_B(\bar m_Q;x)$ does not depend on $c$ and its
explicit expression is not important here.  Thus, only in the
uniform limit ${x, \bar m_Q^2 \rightarrow \infty}$ the heavy quark
propagator may become the free one up to the composition $\bar
D_B(\bar m_Q;x)$, similar to Eqs. (3.28).

 It is instructive to present explicitly a few first terms of
the expansion (3.14) for the $B(x)$ function by substituting it
directly into the general solution (2.3), on account of the known
already expansion for the $A(x)$ function. Omitting all the
tedious algebra one obtains

\begin{equation}
B^2_0(x) = 1 - e^{2(c-x)}, \quad B_2(x) =\Bigl[- x -1 +
e^{2(c-x)}(c-2) \Bigr] B_0^{-1}(x),
\end{equation}
 and all odd terms are zero. In order to reproduce the free quark
propagator case, one has to go again to the limit $x \rightarrow
\infty$ and fixed $c$. Neglecting then the exponentially
suppressed terms and choosing the negative sign in front of the
square root, i.e., $B_0(x) = -1$, one obtains $B_2(x) = x+1$ and
so on. Summing up, one gets the free quark propagator solution
(3.28), indeed. However, even in this case there is a solution
with opposite sign, corresponding apparently to the free heavy
anti-quark propagator. Concluding, let us remind that it is a
general feature of nonlinear systems, like the initial system
(2.2), that the number of independent solutions is not fixed $a \
priori$.

\section{Conclusions}

We have shown that the quark propagator in 2D covariant gauge QCD
reveals several desirable and promising features, so our
conclusions are:

1). The quark propagator has no poles, indeed (quark confinement).

2). It also implies DBCS, i.e., a chiral symmetry is certainly
dynamically (spontaneously) broken for light quarks, while a
chiral symmetry preserving solution is forbidden.

3). The chiral limit physics (i.e., the Goldstone sector) can be
exactly evaluated, since we have found exact solution for the
quark propagator in this case.

4). We develop an analytical formalism, the so-called chiral
perturbation theory at the fundamental quark level, which allows
one to find solution for the quark propagator in powers of the
light quark masses. Each correction satisfies the differential
equation, which can be formally solved exactly.

 5). We develop also an analytical formalism, which allows one to find solution
 for the quark propagator in the inverse powers of the heavy quark masses.
It coincides with the free quark propagator up to terms of order$1
/ \bar m^3_Q$. So this solution automatically possesses the heavy
quark flavor symmetry up to terms of order $1 / \bar m_Q$.

6). We have proved that the exact solution for the $A(x)$ function
coincides with the direct solution of the initial system (2.2),
obtained by using the above-mentioned expansions in the inverse
powers of the heavy quark masses.

7). At the same time, the exact solution (4.7) for the heavy quark
mass function $B(x)$ by no means can be reduced to the free heavy
quark propagator. So it is not coincided with the solution
obtained by the expansions in the inverse powers of the heavy
quark masses.

8). There is no doubt left that by using the expansions in the
inverse powers of the heavy quark masses from the very beginning,
we are loosing some piece of the important information on the
structure of the nonlinear system (2.2) itself. So its direct
solution, obtained by using the straightforward expansions in the
inverse powers of the heavy quark masses, has a particular
character. Evidently, such straightforward solution does not take
into account the response of the vacuum, which determines the
modification of the quark propagator, while the exact solution
(4.7) does take this response into the consideration.

9). The general solution (2.3) does not demonstrate the principal
difference in the analytical structure for light and heavy quarks
propagators. Also at the fundamental quark level the heavy quark
mass limit is not Lorentz covariant. That is why in the case of
heavy quarks we will use the general solution (4.7) rather than
solutions (3.30) and (3.36). To take into account the vacuum's
response is important even for heavy quarks.

10). Our approach makes it possible to calculate physical
observables from first principles. All results will depend only on
the renormalized coupling constant (which has the dimensions of
mass) and the corresponding constant of integration. A physically
well-motivated scale-setting scheme is only needed to fix them.

Our general conclusion is that 2D covariant gauge QCD implies
quark confinement and dynamical breakdown of chiral symmetry
without explicitly involving some extra degrees of freedom. The
only dynamical mechanism responsible for them, which can be
thought of in 2D covariant gauge QCD, is the direct interaction of
massless gluons \cite{7,8}. This interaction is a main dynamical
effect not only in 2D QCD but in 4D QCD as well, i.e., in QCD
itself. However, to directly generalize the quark confinement
mechanism of 2D covariant gauge QCD to 4D QCD is impossible. The
problem is that in former theory the coupling constant, having the
dimensions of mass, plays the role of a mass gap, which was
introduced and discussed by Jaffe and Witten (JW) in Ref.
\cite{12}. In latter theory the coupling constant is
dimensionless, and there is none of the characteristic scales in
QCD Lagrangian. If QCD confines then such a characteristic scale
is very likely to exist, and it is not $\Lambda_{QCD}$, of course,
which can be considered as responsible for the nontrivial PT
dynamics in QCD (scale violation, asymptotic freedom). A possible
way how to introduce a mass gap responsible for the nontrivial NP
dynamics in QCD has been described in Ref. \cite{13}. Its possible
relation to the above-mentiond JW mass gap is also discussed
there.

A financial support from HAS-JINR Scientific Collaboration Fund
(P. Levai) and Hungarian Scientific Fund OTKA T30171 (K. Toth) is
to be acknowledged.

\end{document}